\documentclass[11pt]{article}
\pdfoutput=1

\usepackage[normalem]{ulem}
\usepackage{amsfonts}
\usepackage{longtable}
\usepackage{makeidx}
\usepackage{fullpage}
\usepackage{amsmath}
\usepackage{amssymb}
\usepackage{setspace}
\usepackage{bbm}
\usepackage{dsfont}
\usepackage{graphics}
\usepackage{epsfig}
\usepackage[font=footnotesize,labelfont=bf,justification=centerlast,width=.94\textwidth]{caption}
\usepackage{cite}
 \usepackage{multirow}
\usepackage{array,booktabs}
\usepackage{color}

\usepackage{hyperref}

\hypersetup{
    bookmarks=true,%
    colorlinks,%
    citecolor=blue,%
    filecolor=red,%
    linkcolor=blue,%
    urlcolor=blue
}

\newcommand{\be}{\begin{equation}}
\newcommand{\ee}{\end{equation}}
\newcommand{\bes}{\begin{equation*}}
\newcommand{\ees}{\end{equation*}}
\newcommand{\bea}{\begin{eqnarray}}
\newcommand{\eea}{\end{eqnarray}}
\newcommand{\beas}{\begin{eqnarray*}}
\newcommand{\eeas}{\end{eqnarray*}}

\newcommand{\dd}{\mathrm{d}}

\newcommand{\p}{\partial}

\newcommand{\bt}{\bar{\tau}}
\newcommand{\bl}{\bar{\ell}}
\newcommand{\bmat}{\begin{bmatrix}}
\newcommand{\emat}{\end{bmatrix}}

\newcommand{\RR}{\mathbb{R}}

\newcommand{\ZZ}{\mathbb{Z}}

\def\le{\left}
\def\ri{\right}

\def\t{\tau}
\def\e{\epsilon}

\def\z{\zeta}

\def\le{\left}
\def\ri{\right}

\newcommand{\ben}{\begin{enumerate}}
\newcommand{\een}{\end{enumerate}}

\begin{document}
\numberwithin{equation}{section}
{
\begin{titlepage}
\begin{center}

\hfill \\
\hfill \\
\vskip 0.75in

{\Large \bf Warped Symmetries of the Kerr Black Hole}\\

\vskip 0.4in

{\large Ankit Aggarwal${}^{a,b}$, Alejandra Castro${}^{b}$, and St\'ephane Detournay${}^{a}$}\\

\vskip 0.3in

${}^{a}${\it Physique Mathematique des Interactions Fondamentales, Universite Libre de Bruxelles, Campus Plaine - CP 231, 1050 Bruxelles, Belgium} \vskip .5mm
${}^{b}${\it Institute for Theoretical Physics Amsterdam and Delta Institute for Theoretical Physics, University of Amsterdam, Science Park 904, 1098 XH Amsterdam, The Netherlands} \vskip .5mm

\texttt{ankit.aggarwal@ulb.be, a.castro@uva.nl, sdetourn@ulb.ac.be}

\end{center}

\vskip 0.35in

\begin{center} {\bf ABSTRACT } \end{center}
We propose a set of diffeomorphism that act non-trivially near the horizon of the Kerr black hole. We follow the recent developments of Haco-Hawking-Perry-Strominger to quantify this phase space, with the most substantial difference being our choice of vectors fields.  Our gravitational charges are organized into a Virasoro-Kac-Moody algebra with non-trivial central extensions. We interpret this algebra as arising from a warped conformal field theory. Using the data we can infer from this warped CFT description, we capture  the thermodynamic properties of the Kerr black hole. 

\vfill

\noindent \today

\end{titlepage}
}

\newpage

\tableofcontents

\section{Introduction}

One remarkable success of the holographic principle is the link it provides between two dimensional  conformal field theories, CFT$_2$, and the microscopic origin of black hole thermodynamics.  This first connection was observed in the seminal derivation of the microscopic entropy of five-dimensional extremal BPS black holes in the context of string theory by Strominger and Vafa \cite{Strominger:1996sh}.
Despite the specificity of  the construction,  it provided a framework to explore the robustness of the result beyond its initial scope. Within supersymmetric theories the agreement has been outstanding, allowing for an exploration of both perturbative and non-perturbative effects in quantum gravity; see \cite{Sen:2014aja} for a review on the subject.\footnote{And see \cite{DeHaro:2019gno} for a historical account of these developments in the past two decades.}

Soon after this success, it became clear that the key feature was not supersymmetry, or the specific string theory ingredients in \cite{Strominger:1996sh},  but rather the \textit{conformal symmetry in the near horizon region}.  Conformal invariance is implied by the presence of an AdS factor in the near horizon region, and hence AdS/CFT is responsible for the agreement. For example, many of the above supersymmetric solutions contain an AdS$_3$ factor,  and it was soon evident that part of the above success relied on  AdS$_3$/CFT$_2$. Among other properties, the entropy of these  black holes agrees precisely with the Cardy's growth of states for a CFT$_2$ \cite{Strominger:1997eq,Maldacena:1998bw}.  These developments complemented perfectly the derivation of Brown and Henneaux \cite{Brown:1986nw}, where the symmetry algebra of the so-called large diffeomorphisms forms two copies of a Virasoro algebra. This placed AdS$_3$ gravity, and the black holes within \cite{BTZ,BHTZ}, as a lamppost for our understanding of quantum black holes.

The role of 2d conformal symmetry might not be restricted to extremal black holes with a decoupled near-horizon AdS$_3$ region, with early  indications already present in \cite{Maldacena:1997ih,Cvetic:1997xv,Cvetic:1997ap}. One important observation in this direction was to highlight a hidden conformal symmetry of the Klein-Gordon operator on the Kerr background, which is made manifest when considering a ``near'' region of phase space rather than spacetime \cite{Castro:2010fd}. A second observation is the explicit identification of two sets of Virasoro diffeomorphisms acting on the horizon   \cite{Haco:2018ske}. The central extensions derived there fill an important gap in this program that address the microscopic origin of the entropy of a generic Kerr black hole with mass $M$ and angular momentum $J$. More concretely, the Cardy formula in a CFT$_2$
\be \label{CardyEntropy}
S_{\rm Cardy} = \frac{\pi^2}{3} c \, (T_L + T_R)~. 
\ee
with left and right-moving temperatures in \cite{Castro:2010fd}, and central charge $c=12J$ obtained in \cite{Haco:2018ske}, reproduces the Bekenstein-Hawking area law of the Kerr black hole. 

The question we would like to ask here is whether the arguments used in \cite{Castro:2010fd, Haco:2018ske} that point at a description of generic Kerr black holes in terms of a CFT$_2$ are unique. In recent years indeed, alternative holographic scenarios have started to emerge in which the geometries no longer exhibit a local  $SL(2,\RR) \times SL(2,\RR)$ symmetry, and where the dual field theories differ from a traditional CFT$_2$. One such scenarios involves a new type of field theories, called Warped Conformal Field Theories (WCFT) \cite{Hofman:2011zj, Detournay:2012pc}. We will argue that a description of generic Kerr black holes in terms of a WCFT might appear as natural as a CFT$_2$ one and, in particular, that it allows to reproduce the Bekenstein-Hawking entropy along the lines of \cite{Haco:2018ske}.


WCFTs are two-dimensional non-relativistic field theories, with an $SL(2,\RR) \times U(1)$ global symmetry that gets extended into an infinite-dimensional Virasoro-Kac-Moody algebra \cite{Hofman:2011zj, Detournay:2012pc}. Many of their unique field theoretic properties have been uncovered in recent years \cite{Hofman:2014loa, Castro:2015uaa,  Detournay:2015ysa, Castro:2015csg, Song:2016gtd, Song:2016pwx, Song:2017czq, Jensen:2017tnb, Apolo:2018eky, Apolo:2018oqv, Song:2019txa, Chen:2019xpb}. In a holographic context, their natural counterparts are gravitational theories that admit Warped AdS$_3$ (WAdS$_3$) spacetimes as solutions \cite{Israel:2004vv, Rooman:1998xf, Moussa:2003fc, Anninos:2008fx}. 
For several instances of WAdS$_3$ one can establish that the asymptotic symmetries form a Virasoro-Kac-Moody algebra, which is an extension of their local $SL(2,\RR) \times U(1)$ Killing symmetries \cite{Compere:2007in, Compere:2008cv, Compere:2009zj, Anninos:2010pm, Henneaux:2011hv, Blagojevic:2009ek, Anninos:2011vd}, thus precisely matching the defining symmetries of a WCFT. Remarkably, WCFTs exhibit modular properties that allow to derive a Cardy-type formula for the asymptotic degeneracy of states, which is found to match the entropy of the corresponding WAdS$_3$ black holes \cite{Detournay:2012pc, Donnay:2015iia,Giribet:2015lfa, Afshar:2015wjm, Detournay:2016gao, Zwikel:2016smm, Setare:2017nlu, Azeyanagi:2018har},
in the spirit of \cite{Strominger:1997eq}. This is taken as an indication that WCFTs could be relevant to give a microstate description of certain classes of black holes in lower dimensions. In this work, we will investigate whether WCFTs could also play a role to understand higher dimensional black holes, in particular non-extremal Kerr black holes in 4 dimensions.\footnote{WAdS$_3$ spaces  famously appear in the near-horizon geometry of extremal four-dimensional Kerr black holes \cite{Bardeen:1999px} central to the Kerr/CFT correspondence \cite{Guica:2008mu}, which makes this possible holographic description worth exploring.}

Our paper is organized as follows. We will start in Sec.\,\ref{sec:bhmonodromy} by reviewing the analytic properties of the Klein-Gordon operator on the Kerr background. This operator has two regular singular points with non-trivial monodromies, located at the inner and outer horizon of the Kerr black hole. As shown in \cite{Castro:2013kea},  these monodromies  give us a natural basis of energy eigenstates that are crucial for the identification of vector fields we will use Sec.\,\ref{sec:virkm} and the identification of $T_{L,R}$ in \eqref{CardyEntropy}. Although this choice of basis is equivalent to the one used in \cite{Haco:2018ske}, the advantage of phrasing this choice in terms of monodromies is that it does not rely on a low frequency limit of scattering amplitudes: it is an exact feature about the analyticity of an eternal black hole. 

Sec.\,\ref{sec:virkm} contains our main result: we will propose a set of vector fields, whose algebra fits the symmetries of a WCFT. As in \cite{Haco:2018ske}, we use Wald-Zoupas formalism to define covariant charges and evaluate the central extensions of the gravitational charges associated to these vectors. We find a non-trivial central extension for the Virasoro commutator and the Kac-Moody current. One interesting feature of our analysis is that the integrals involved in these terms only receive non-trivial contributions from the future horizon.

In Sec.\,\ref{sec:thermo} we discuss how one could use the thermodynamic properties of a WCFT to account for the Bekenstein-Hawking entropy. Using the basis of monodromies in Sec.\,\ref{sec:bhmonodromy} and the central extensions in Sec.\,\ref{sec:virkm}, we are able to account for the entropy of the Kerr black hole using the asymptotic formula for the growth of states in a WCFT. This is another piece of evidence that advances a different holographic description of the Kerr black hole. 

We end our work with a discussion that includes future directions and weaknesses in this program that are worth mentioning. In App.\,\ref{app:wcft} we provide some introductory material on WCFT based on \cite{Detournay:2012pc,Castro:2015uaa}, and in App.\,\ref{app:cardy} we review the derivation of the Cardy formula using the analogous approach as done for WCFT. Finally in App.\,\ref{app:wave} we discuss how warped symmetries could be hidden in the low frequency limit of a scalar wave equation.

\section{Black hole monodromy}\label{sec:bhmonodromy}

The metric for the Kerr black hole,  in Boyer-Lindquist coordinates, is given by
\be\label{app:geom}
ds^2={\rho\over {\Delta}}dr^2- {{\Delta}\over \rho}\left(dt -{a}\sin^2\theta\, d\phi\right)^2+ \rho \, d\theta^2+{ \sin^2\theta \over \rho} \left((r^2+{a}^2)\,d\phi-{{a}}\,dt\right)^2~,
\ee
where $a$ is a constant parameter that controls the rotation, and $\rho = r^2+{a}^2\cos^2\theta$. The mass of the black hole $M$ enters the metric via 
\be
\Delta = r^2 +{a}^2-2M r~.
\ee
The inner and outer horizons are located at
\be\label{rpm1}
{r}_\pm = M \pm \sqrt{M^2-{a}^2}~,
\ee
which corresponds to the zeroes of $\Delta$. The angular momentum of the black hole is $J=Ma$.

As we discuss physical observables in Kerr, one of our agenda points will be to exploit analytic properties of the black hole background.  
To motivate our later choices, we will review how analyticity, represented by monodromy data, enters in the structure of the Klein-Gordon equation \cite{Castro:2013kea,Castro:2013lba}. The starting point is to consider a massless scalar field $\psi$ in the Kerr background, i.e.
\be\label{bb:KG}
{1\over \sqrt{-g} } \partial_\mu\left(\sqrt{-g}g^{\mu\nu}\partial_\nu\psi\right)=0~.
\ee
Expanding in eigenmodes 
\be\label{bb:eigen}
\psi(t,r,\theta, \phi) = e^{-i\omega t+im\phi}R(r)S(\theta)~,
\ee
the Klein-Gordon operator \eqref{bb:KG} becomes separable. The spheroidal equation is
\be\label{bb:angeqn}
\left[ {1\over \sin\theta}\partial_\theta
\left(\sin\theta\partial_\theta\right)-{m^2\over\sin^2\theta}+\omega^2 a^2\cos^2\theta \right] S(\theta)=-K_\ell  S(\theta)~,
\ee
for eigenvalue $K_\ell(a\omega)$; to leading order $K_\ell(a\omega)= \ell(\ell+1)+O(a\omega)^2$. And  the radial equation is given by
\begin{align}\label{bb:10}
\Bigg[
\partial_r (r-r_-)(r-r_+) \partial_r
+{(\omega-\Omega_+ m )^2\over 4\kappa_+^2}{(r_+-r_-) \over (r-r_+)}   -{(\omega-\Omega_- m )^2\over 4\kappa_-^2}{ (r_+-r_-)\over (r-r_-)} \cr  +(r^2+2M(r+2M) )\omega^2 \Bigg] R(r) =K_\ell  R(r)~ ,
\end{align}
where
\be\label{eq:HK}
\kappa_{\pm}= {r_+-r_-\over 4M r_\pm}~,\quad \Omega_\pm ={a\over 2Mr_{\pm}}~,
\ee
are the surface gravity and angular velocity evaluated at the inner ($r=r_-$) and outer horizon ($r=r_+$).

The global properties of solutions to \eqref{bb:10} are as follows.
Equation \eqref{bb:10} has two regular singular points at $r=r_+$ and $r=r_-$, which means that the solutions to \eqref{bb:10} have branch cuts at these points.
For instance, around  $r=r_+$ we have two linearly independent solutions
\be
\label{n1kerr}
R_+^{out}(r)  = (r-r_+)^{i\alpha_+} \big( 1 + O(r-r_+)\big)~,\qquad
R_+^{in}(r)  = (r-r_+)^{-i\alpha_+} \big( 1 + O(r-r_+)\big)~,
\ee
where  
\be\label{bb:residue+}
\alpha_+ = {(\omega-\Omega_+ m )\over 2\kappa_+}~.
\ee
This is a convergent series expansion for $R(r)$ when $|r-r_+| < |r_--r_+|$.
Near the the inner horizon, $r=r_-$, we have a similar expansion, where $\alpha_+$ is replaced by
\be\label{bb:residue-}
\alpha_- = {(\omega-\Omega_- m )\over 2\kappa_-}~.\ee
Here $\alpha_\pm$ control the analyticity of the solutions: they are parameters that measure the lack of meromorphicity   in $R(r)$ as one transports a solution in the complex $r$-plane around one of the singular points. Note that the monodromy eigenvalues $\alpha_\pm$ in \eqref{bb:residue+}-\eqref{bb:residue-} are exact statements about the differential equation; it is not an artefact of a low frequency regime or any other limit.  

There is also a singular point at $r=\infty$, and the series expansions reads
\be
\label{n2kerr}
R^{out}_\infty(r) \sim e^{i \omega r} r^{ i \lambda-1} \big(1+ O(r^{-1})\big)~,\qquad
R^{in}_\infty(r) \sim e^{-i \omega r} r^{- i \lambda-1} \big(1+ O(r^{-1})\big)~,
\ee
 with
\be
 \lambda =  2M \omega ~.
\ee
Unlike the singular points at $r=r_\pm$, however,  the singular point at $r=\infty$ is irregular. 
This means that the series \eqref{n2kerr} is asymptotic rather than convergent.  The series expansion for $R(r)$ appearing in \eqref{n2kerr} is therefore referred to as a formal solution to the wave equation, as opposed a true solution defined by an analytic continuation of the series expansion \eqref{n1kerr} to the whole complex $r$-plane. The true solution around this irregular point has a monodromy, $\alpha_{\rm irr}$, which can be computed perturbatively in $\omega$ \cite{Mano:1996vt,Mano:1996gn,Castro:2013lba}. Its leading behaviour is $i\alpha_{\rm irr}=\ell + O(\omega^2)$ with $\ell$ the spherical harmonic that controls \eqref{bb:angeqn}.

The monodromy parameters $\alpha_\pm$ connect in an elegant manner the scattering coefficients to the analytic properties of the geometry, see e.g. \cite{Castro:2013kea,Castro:2013lba,Novaes:2014lha,daCunha:2015ana}. Their use here will be as energy eigenstates that will motivate our choice of vector fields in Sec.\,\ref{sec:virkm}, and impact the holographic interpretation of the thermodynamics of Kerr in Sec.\,\ref{sec:thermo}.   The canonical choice of eigenstates is to consider functions of the eigenvalues $(\omega, m)$ of the operators $(i\partial_t, -i\partial_\phi)$, as done in \eqref{bb:eigen}. Following \cite{Castro:2013kea}, we will consider instead the linear combinations of the monodromies:
\bea\label{eq:omegaLR}
 \omega_L &:=& \alpha_+ - \alpha_-~,\cr
 \omega_R&:=& \alpha_++ \alpha_-~.
\eea
To the energies $\omega_{R,L}$ we will assign conjugate variables $t^{\pm}$, with $(\omega_{L},\omega_R)$ eigenvalues of $\left(i\partial_{t^-},i\partial_{t^+}\right)$.  Thus, 
\be
e^{-i\omega t +i m \phi}=e^{-i{\omega_L} t^- -i{\omega_R} t^+ }~.
\ee
Using the explicit form of the monodromies for Kerr \eqref{bb:residue+}--\eqref{bb:residue-}, we find that 
\bea\label{monbasis}
t^+ =2\pi T_R \,\phi~,\qquad 
t^- = {1\over 2M}t - 2\pi T_L\, \phi~,
\eea
where $(t,\phi)$ are Boyer--Lindquist coordinates, and 
\be\label{eq:temp}
T_L= {r_++r_-\over 4\pi a}~,\qquad T_R= {r_+-r_-\over 4\pi a}~.
\ee
In position space, we will rewrite \eqref{eq:omegaLR}
\bea\label{eq:v2}
 H_0={i \over 2\pi T_R}\partial_\phi +2iM{T_L \over T_R}\partial_t~,  \qquad \bar H_0=-2iM\partial_t~. 
\eea
It is important to note that these vectors are the same as those used to exhibit the hidden conformal symmetry of Kerr  \cite{Castro:2010fd}, and used in \cite{Haco:2018ske}. In our subsequent derivations, a difference worth highlighting,  is that we are using the monodromy basis to select \eqref{eq:v2}.  Moreover we will  not interpret \eqref{eq:v2} as a basis for CFT$_2$ energy eigenstates; instead we explore an alternative interpretation in terms of a WCFT description.

\section{Warped symmetries in the Kerr geometry}\label{sec:virkm}

In this section we will introduce a the set of diffeomorphisms in the near horizon geometry of the Kerr black hole, and study the linearized charges associated to them. This section follows closely the proposal in \cite{Haco:2018ske}, where the most significant deviation comes from our choice of vector fields.

As in \cite{Castro:2010fd,Haco:2018ske} it is convenient to adapt ``conformal'' coordinates
$(w^\pm,y)$ defined in terms of $(t,r, \phi)$ by
\begin{align}\label{ConfCoord}
w^+ &=\sqrt{r-r_+\over r-r_-}\, e^{2\pi T_R \phi}~,\cr 
w^- &=\sqrt{r-r_+\over r-r_-}\,e^{2\pi T_L \phi-{t \over 2M}}~,\cr
y &=\sqrt{r_+-r_-\over r-r_-}\,e^{\pi (T_L+T_R) \phi-{t \over 4M}}~.
\end{align}
A motivation to introduce this coordinate system, is that the local vector fields in \eqref{eq:v2} are now independent of the Kerr black hole parameters; and in particular
%
%
\be\label{eq:h2}
H_0=i(w^+\p_++{1 \over 2}y\p_y)~,\qquad \bar H_0=i(w^-\p_-+{1 \over 2}y\p_y)~. 
\ee
An important feature is that under azimuthal identification $\phi \sim \phi+2\pi$ one finds
\be\label{idn} 
w^+ \sim e^{4\pi^2 T_R}w^+~, ~~ 
w^- \sim e^{4\pi^2 T_L}w^-~, ~~
y\sim e^{2\pi^2 (T_R+T_L)}y~.
\ee

With the intention of promoting warped symmetries as a holographic description of the Kerr black hole, we will introduce a set of suitably chosen vector fields. These are 
\begin{align} \label{virU1}
\zeta(\epsilon)&= \epsilon(w^+)\, \partial_++{1\over 2}\partial_+\epsilon(w^+) \,y\partial_y~,\cr
 p(\hat\e)&=  ~\hat\e(w^+)\, (w^- \p_- +{y\over 2} \p_y) ~ ,
\end{align}
 where $\epsilon$ and $\hat \e$ are arbitrary functions of $w^+$. We restrict these functions such that the vector fields (\ref{virU1}) are periodic under \eqref{idn}; a Fourier decomposition achieving this is
 \be\label{csf}
 \e_n={2 \pi T_R}(w^+)^{1+{in \over 2 \pi T_R}}~,\qquad \hat\e_{n'}=(w^+)^{{in' \over 2 \pi T_R}}~,
 \ee
with $n, n' \in \ZZ$, and we define $\zeta_n\equiv \zeta(\e_n)$ and $p_n\equiv p(\hat \e_n)$. With this choice, the Lie bracket of the above vector fields reads
\begin{align} \label{VKM}
i [\zeta_m, \zeta_n ]&= (m-n) \zeta_{m+n}~, \cr
i [\zeta_m, p_n]&= n\, p_{m+n}~, \cr 
i [p_m, p_n]&= 0~,
\end{align}
 which is a Virasoro-KacMoody (VKM) algebra without any central extension.
 In comparison to the vector field in \eqref{eq:h2}, we have
\be
\zeta(\e_0)= -i 2\pi T_R H_0 ~,\quad p(\hat \e_0)= i\bar H_0~,
\ee
i.e. our zero modes coincide with the monodromy basis.




\subsection{Central extensions}

The subsequent analysis will be done using conformal coordinates, and for that purpose it is useful to record some of their properties. 
The past horizon, which in the Boyer Lindquist systems is at $r=r_+, ~ t\in (-\infty,~ 0)  $, maps to $w^+=0$; the future horizon, located at $r=r_+, ~ t\in ( 0, ~ \infty) $, maps to $w^-=0$.  The bifurcation surface $\Sigma_{\rm bif}$ is therefore at $w^+=w^-=0$, and  around this  surface the metric \eqref{app:geom} becomes 
  \begin{align} \label{conformal metric}
   ds^2&= {4 \rho_+^2 \over y^2} d w^+ dw^-
+  {16 J^2 \sin^2\theta \over y^2 \rho_+^2} dy^2 +\rho_+^2 d\theta^2 
- {2w^+ (8\pi J)^2 T_R(T_R+T_L)  \over y^3 \rho_+^2} dw^- dy \cr  
&\quad+ {8 w^- \over y^3 \rho_+^2} \big(- (4\pi J)^2T_L(T_R+T_L) + (4 J^2 + 4\pi J a^2 (T_R+T_L)  + a^2 \rho_+^2) \sin^2\theta \big) dw^+ dy \cr
&\quad+ \cdots~, 
\end{align}  
where corrections are at least second order in $(w^+,w^-)$.

We can associate to the vector fields \eqref{virU1} covariant charges; these charges implement the symmetries on a phase space via the Dirac bracket. 
The linearized charges we will be studying are
\be\label{eq:defcharges}
\delta \mathcal{Q} = \delta \mathcal{Q}_{\rm IW} + \delta\mathcal{Q}_{\rm WZ}~.
\ee
The first term is the Iyer-Wald charge, which reads
\be\label{eq:iwc}
 \delta \mathcal{Q}_{\rm IW} (\chi,h ; g)= {1\over 16\pi} \int_{\partial\Sigma} \star F_{\rm IW}~.
\ee
The surface of integration is the bifurcation surface, $\partial \Sigma=\Sigma_{\rm bif}$. The input in this definition includes a  vector field $\chi$, which we will take to be \eqref{virU1}, a metric perturbation $h_{\mu\nu}$, and the background metric $g_{\mu\nu}$ in \eqref{conformal metric}.  The integrand is given by
\be
(F_{\rm IW})_{\mu\nu}={1\over2}\nabla_\mu \chi_\nu \, h +\nabla_\mu h^{\lambda}_{~\nu}\, \chi_{\lambda}+\nabla_\lambda \chi_{\mu} \, h^{\lambda}_{~\nu}- \nabla_\mu h \,\chi_{\nu} - (\mu\leftrightarrow \nu)~.
\ee
The second term in \eqref{eq:defcharges} is the Wald-Zoupas counterterm \cite{Wald:1999wa}, introduced in \cite{Haco:2018ske} to comply to some consistency conditions which we will highlight below. Its definition is
\be\label{eq:wzc}
 \delta \mathcal{Q}_{\rm WZ} (\chi,h ; g) = {1\over 16\pi }\int_{\partial \Sigma} \iota_{\chi}(\star X) ~, \qquad X =2 h_{\mu}^{~\nu} \Omega_\nu \dd x^\mu~,
\ee
where $\Omega_\mu$ is defined as
\be
\Omega_\mu = q^\lambda_{~\mu}n^\nu \nabla_{\lambda} l_\nu~.
\ee
The vectors $n^\mu$ and $l^\mu$ are null, normal to $\Sigma_{\rm bif}$, and $n\cdot l =-1$; $q_{\mu\nu}=g_{\mu\nu} +n_{\mu}l_{\nu}+n_{\nu}l_{\mu}$ is the induced metric on $\Sigma_{\rm bif}$. In addition we demand that $n^\mu$ and $l^\mu$ are single valued under \eqref{idn}, which fixes them up to a rescaling.

We are interested in quantifying if the algebra of charges associated to \eqref{VKM} admits a central extension. More explicitly, we will define 
\be\label{eq:covcharge}
\delta L_n \equiv  \delta \mathcal{Q} (\zeta_n, h;g) ~,\qquad \delta P_n \equiv  \delta \mathcal{Q} (p_n, h;g)~,
\ee
for the vectors in \eqref{virU1} and \eqref{csf}. Assuming that these charges are integrable and that the Dirac bracket is well defined, there are 3 possible central extensions. For the Virasoro sector we have 
\be \label{cterm}
[L_n, L_m]= (m-n) L_{m+n} +  K_{m,n}~, \qquad K_{m,n}=\delta \mathcal{Q}(\z_n,\mathcal L_{\z_m}g;g)~,
 \ee
and the Kac-Moody generators
\be\label{eq:kmterm}
[P_n, P_m]=  k_{m,n}~, \qquad k_{m,n}=\delta \mathcal{Q}(p_n,\mathcal L_{p_m}g;g)~.
\ee
And finally there is as well the possibility of a mixed central term
\be
[L_n, P_m]= m P_{n+m} + \mathfrak{K}_{m,n}~, \qquad \mathfrak{K}_{m,n}=\delta \mathcal{Q}(\zeta_n,\mathcal L_{p_m}g;g)~.
\ee
The evaluation of these central terms involves some subtleties which we now turn to.

As it was observed in  \cite{Haco:2018ske}, the leading singularities in the integrand of $\delta{\cal Q}$ near $\Sigma_{\rm bif}$ are at most simple poles. For example, in evaluating $K_{m,n}$ there is a non-zero contribution from   the $F^{-y}_{\rm IW}$ component, i.e., from the constant $y$ cross-section of the future horizon close to the bifurcation surface. This term contains a simple pole in $w^+$, and the relevant integral to evaluate is
\be \label{orthogonality}
 \int^{w_0^+ e^{4 \pi^2 T_R}}_{w_0^+}{d w^+\over w^+} =4\pi^2 T_R~.
\ee 
The limits of $w^+$ are governed by the identification $w^+\sim w^+ e^{4 \pi^2 T_R}$, and $w_0^+$ is a reference point near $w^+\to 0$. The Wald-Zoupas term in \eqref{eq:wzc} has the same type of singular behaviour. The evaluation of $K_{m,n}$ therefore receives contributions from 
\begin{align}\label{eq:k1}
\delta \mathcal{Q}_{\rm IW}(\z_n,\mathcal L_{\z_m}g;g)&= 2J {T_R\over T_L+T_R}\le((-1+{M^2\over a^2})m + m^3\ri)\delta_{n,-m}~,\cr
\delta \mathcal{Q}_{\rm WZ}(\z_n,\mathcal L_{\z_m}g;g)&= J {T_L-T_R\over T_L+T_R}\le((-1+{M^2\over a^2})m + m^3\ri)\delta_{n,-m}~,
\end{align}
where we used \eqref{orthogonality} to evaluate the integral over $w^+$. The linear term in $m$ can be reabsorbed in the zero mode of the generators, and will be ignored in the following. Adding up the $m^3$ contributions gives
\be\label{eq:ctot}
K_{m,n}= J\,m^3\, \delta_{n,-m} ~,
\ee
which reproduces the right-movers in \cite{Haco:2018ske}. As noted there, individually the terms in \eqref{eq:k1} are problematic since they depend on the mass of the black hole; removing this dependence is the main  motivation to introduce the Wald-Zoupas counterterm. 


%
For the two remaining central extensions, the steps are exactly the same: the only non-zero contribution  in the Iyer-Wald term \eqref{eq:iwc} and Wald-Zoupas counterterm in \eqref{eq:wzc} comes from the constant $y$ cross-section of the future horizon close to the bifurcation surface. Thus, we are left to evaluate an integral over $\theta$ and $w^+$, where the integral over $w^+$ is of the form \eqref{orthogonality}. This is an important difference relative to \cite{Haco:2018ske}: all of our central extension have their pole in the future horizon. We will comment more on this feature in the discussion.

Carrying out the appropriate integrals, the central extension $k_{m,n}$ in \eqref{eq:kmterm} receives the two non-trivial contributions which are
\begin{align}
\delta \mathcal{Q}_{\rm IW}(p_n,\mathcal L_{p_m}g;g)= - 2J {T_L\over T_L+T_R} m\,\delta_{n,-m}~,\cr
\delta \mathcal{Q}_{\rm WZ}(p_n,\mathcal L_{p_m}g;g)= J {T_L-T_R\over T_L+T_R}m\,\delta_{n,-m}~,
\end{align}
where these contribution are due to the integrals over the future horizon and we used \eqref{orthogonality}. And as in \eqref{eq:ctot}, the addition of these terms gives a mass independent result; we find
\be
k_{m,n}=-J\, m\,\delta_{n,-m}~,
\ee
Finally, our proposed set of vector fields could also allow for a mixed central extension $\mathfrak{K}_{m,n}$. In this case have
\begin{align}
\delta \mathcal{Q}_{\rm IW}(\zeta_n,\mathcal L_{p_m}g;g)=i J {T_R-T_L\over T_L+T_R}\bigg( m {i\sqrt{M^2-a^2}\over a}+m^2\bigg)\,\delta_{n,-m} ~,\cr
\delta \mathcal{Q}_{\rm WZ}(\zeta_n,\mathcal L_{p_m}g;g)=-i J {T_R-T_L\over T_L+T_R}\bigg( m {i\sqrt{M^2-a^2}\over a}+m^2\bigg)\,\delta_{n,-m} ~,
\end{align}
which adds  to zero, and hence $\mathfrak{K}_{m,n}=0$. It is interesting to note that Wald-Zoupas counterterm here served the additional purpose of eliminating the mixed central extension for our algebra.

Gathering all the results from evaluating the central extensions, the algebra of covariant charges \eqref{eq:covcharge} associated to the diffeomeorphisms \eqref{virU1}-\eqref{csf} seems to obey\footnote{We emphasise that at this stage we have not evaluated the charges explicitly; we are assuming that they obey suitable Dirac brackets. }

\begin{align}
[L_n, L_m]&= (m-n) L_{m+n} + \frac{c}{12}m^3\delta_{n,-m}~,\cr
[L_n, P_m]&= m P_{n+m}~,\cr
[P_n, P_m]&=k \frac{m}{2}\delta_{n,-m}~,
\end{align}
with
\be \label{CentralExtensions}
c =12 J ~,\qquad k=-2J~.
\ee
This derivation supports the proposal that the Kerr black hole could be described holographically in terms of a WCFT. The non-trivial central extensions for the Virasoro-KacMoody algebra depend on the angular momentum of the black hole. 

\section{Black hole thermodynamics from warped symmetries}\label{sec:thermo}

In this section we will show how the above information can be used to interpret the entropy of the Kerr black hole 
\be \label{KerrEntropy}
S_{\rm BH} = {A_H\over 4}=2 \pi (M^2 + \sqrt{M^4 - J^2})~,
\ee
as the entropy of a thermal state in a WCFT.

The change of coordinates \eqref{ConfCoord} can be rewritten, at fixed $r \ne r_+$, and up to an $r$-dependent rescaling as
\be \label{Rindler}
  w^\pm = e^{\pm t^\pm}~,
\ee
with $t^\pm$ precisely the coordinates in the monodromy basis \eqref{monbasis}.
Since $\phi$ is $2\pi$-periodic, the corresponding identifications on $t^\pm$ make  \eqref{Rindler} the well-known relation between Minkowski ($w^\pm$) and Rindler ($t^\pm$) coordinates. In the Minkowski vacuum, observers at a fixed position in Rindler coordinates will detect a thermal radiation.

Now, the orginal Kerr coordinates are subject to both spatial {\itshape and} thermal identifications, respectively
\be \label{KerrId}
 (t,\phi) \sim (t,\phi +2 \pi) \sim (t + i \beta, \phi + \theta)~,
 \ee
 with $\beta = {2\pi\over \kappa_+}$ the inverse Hawking temperature, and $\theta = i \beta \Omega_+$  the angular potential as defined in \eqref{eq:HK}.
In terms of \eqref{monbasis}, these identifications read as
\be\label{eq:xx1}
(t^+,t^-)\sim  (t^+ + 4 \pi^2 T_R,t^- - 4 \pi^2 T_L)  \sim  (t^+ + 2 \pi i,t^- + 2 \pi i)~.
\ee
The monodromy analysis therefore suggests that in the natural field theory coordinates ($t^\pm$), \emph{thermal and spatial cycles are swapped compared to the original Kerr periodicities}. This is the case for both the CFT description (which is reviewed in App.\,\ref{app:cardy}) and the WCFT one, which we now turn to.
 
Consider a WCFT defined on a generic torus with spatial and thermal identifications
\be \label{tpm}
(t^+,t^-)\sim(t^+-2\pi \ell, t^-+2\pi \bar{\ell})\sim(t^+-2\pi \tau, t^-+2\pi \bar{ \tau})~.
\ee
The entropy in the canonical ensemble, which we review in App.\,\ref{app:wcft},  is given by \eqref{scan} and it reads
\be \label{scan2}
S_{\bl|\ell}(\bt|\t)=2\pi i {z\over \hat t}\hat{P}_0^{\rm vac}+{4 \pi i\over \hat t} \hat{L}_0^{\rm vac}~,
 \ee
with $\hat{P}_0^{\rm vac}, \hat{L}_0^{\rm vac}$ are the vacuum values of the zero modes, and
\be
z=\bt-{\bl\t\over\ell}~, ~ \ {\hat t}={\t\over \ell}~.
\ee

Comparing \eqref{eq:xx1} and \eqref{tpm}, one obtains
\be \label{period}
\ell = -2 \pi T_R~, \quad \bl = -2 \pi T_L~, \quad \tau = -i~, \quad \bt = i~,
\ee
and hence the entropy is given by
\be \label{SWCFT}
 S = 4 \pi^2 i (T_L + T_R) \,  \hat{P}_0^{\rm vac} + 8 \pi^2 T_R \,  \hat{L}_0^{\rm vac}~.
\ee
We point out that here $T_L$ and $T_R$ are defined via the identifications in \eqref{eq:xx1}; in a WCFT they do not have an interpretation in terms of left and right moving temperatures.

To compare expression (\ref{SWCFT}) to the Bekenstein-Hawking entropy of Kerr in \eqref{KerrEntropy}, one needs to identify $\hat{P}_0^{\rm vac}$ and $\hat{L}_0^{\rm vac}$. First note that these two quantities are not independent \cite{Detournay:2012pc}, and related through:
\be\label{eq:vac1}
  \hat{L}_0^{\rm vac} = - \frac{c}{24} + \frac{(\hat{P}_0^{\rm vac})^2}{k}~,
\ee
where $c$ and $k$ are the WCFT central extensions. Furthermore, in a WCFT $\hat{P}_0^{\rm vac}$ is not fixed by the symmetries alone\cite{Detournay:2012pc}, unlike what happens in unitary 2d CFTs.  To fix this ambiguity we will borrow from other instances of WCFT in holography. A consistent pattern in holographic setups, such as \cite{Compere:2007in, Compere:2008cv, Compere:2008cw, Compere:2009zj, Tonni:2010gb, Detournay:2012pc, DetournayGuica2013, Donnay:2015iia},  is
 \be \label{L0vac}
  \hat{L}_0^{\rm vac} = 0~,
  \ee
where the common thread is the absence of a gravitational anomaly in the bulk theory, such as a gravitational  Chern-Simons term.\footnote{Gravitational anomalies in local WCFTs have also been investigated in \cite{Jensen:2017tnb}.} Note that this pattern renders an imaginary value for $\hat{P}_0^{\rm vac}$ in holography. 
%
It is natural to assume that (\ref{L0vac}) holds for our circumstances too!  Using the values of the central extensions in \eqref{CentralExtensions} and \eqref{L0vac} in \eqref{eq:vac1} gives
\be \label{P0vac}
   (\hat{P}_0^{\rm vac})^2 = - J^2~,
\ee 
and the entropy \eqref{SWCFT} is simply
\be \label{SWCFT1}
 S =  {4 \pi^2 } |J|    (T_L + T_R) ~.
 \ee
 Plugging \eqref{eq:temp}, and \eqref{rpm1}, one obtains that \eqref{SWCFT} exactly agrees with \eqref{KerrEntropy}.

\section{Discussion}

We found evidence that a WCFT could be a suitable holographic description of the Kerr black hole. Many conceptual pieces are missing in this description and several aspects of this proposal are mysterious to us. In the following we will discuss future directions that could address these  issues. 

\begin{enumerate}

\item 
One notable aspect of our computation is that the contribution to the central extensions comes only from a component of the future horizon that is near the bifurcation surface at constant $y$.\footnote{In contrast, the central extensions in \cite{Haco:2018ske} have contributions from the future horizon for their right moving vector fields, and past horizon for the left movers.} This hints at the possibility that these symmetries could be used to understand more dynamical situations, like black holes formed by collapse. However, it still remains to be seen whether the charges also follow a similar pattern. As a first step in this direction, we have tried to compute the charges for the Kerr background. Here, we found that the charges receive a finite contribution from the future horizon. However, the other two surfaces cause problems. The contributions from these surfaces are ill-defined unless $\delta r_+=0$, i.e., the horizon radius is kept fixed in the phase space. This is not what we have been using in the central extension computations, where we fixed the angular momentum $J$. It is important to clarify this issue and better understand the phase space.

To get some insight in this matter, we did similar computations in three dimensions for the BTZ black hole in Einstein gravity (and TMG) where the phase space is better understood. We used the natural analogs of the conformal metric \eqref{ConfCoord} and vector fields  \eqref{virU1} adapted to the 3D black hole. After a computation almost identical to the one for Kerr,\footnote{For BTZ we only need to evaluate the Iyer-Wald charge in \eqref{eq:iwc} to reproduce, for example, the Brown-Henneaux central charge \cite{Brown:1986nw}. The Wald-Zoupas counterterm \eqref{eq:wzc} gives a finite contribution to the central extension that  depends on the charges of BTZ. The comments in this paragraph are based on using only \eqref{eq:iwc} in the definition of covariant charges.} we get the CSS central extensions \cite{Compere:2013bya}, and their TMG counterpart. For the charges of the BTZ background, we again find that the future horizon gives finite contributions. The other surfaces contribute pathologically unless $\delta r_+=0$, as for the Kerr background. A similar study done for Warped BTZ in TMG also reproduces the correct central extensions. All these studies in 3d reinforce our confidence for the Kerr computation presented in this work. However, the important issues regarding the charges and phase space for generic Kerr still remain, and are left for future investigations. 

In this regard, it would be interesting to discuss potential ambiguities in the definition of the charges. As proposed in \cite{Haco:2018ske} and as we discussed in this paper, a Wald-Zoupas boundary counterterm had to be added to make the central charge state-independent. It is not known whether the term used here is unique, or if there exists an alternative term that would make both the central charges state-independent {\itshape and} the near-horizon charges well-defined for example. It would be clearly desirable to find a first principle derivation fixing this boundary term, for instance by requiring to have a well-defined variational problem.


\item Our work illustrates that there is ambiguity on defining horizon symmetries for non-extremal black holes.
Earlier related works by Carlip studying horizon symmetries for generic black holes exhibit the same ambiguity: depending on the choice of boundary conditions at the horizon, the symmetries can be either conformal \cite{Carlip:1994gy,  Carlip:1998wz, Carlip:2011vr, Carlip:2012ff} or BMS$_3$ \cite{Carlip:2017xne}, and both seem able to account for the black hole entropy. It is also worth mentioning that various proposals for boundary conditions at non-extremal horizons in three and four dimensions have appeared in recent years \cite{Donnay:2015abr, Afshar:2015wjm, Afshar:2016wfy, Donnay:2016ejv, Afshar:2016kjj, Grumiller:2017otl, Gonzalez:2017sfq, Donnay:2019zif, Donnay:2019jiz, Grumiller:2019fmp}. The ambiguity in the choice of boundary conditions is reminiscent of the archetypical AdS$_3$ gravity setting: besides the Brown-Henneaux boundary conditions, a handful of alternative boundary conditions have appeared in recent years with symmetries differing from those of a pure CFT$_2$ \cite{Compere:2013bya, Troessaert:2013fma, Avery:2013dja, Grumiller:2016pq, Perez:2016vqo, Donnay:2015abr,  Afshar:2015wjm, Afshar:2016wfy}. In the case at hand, it would be interesting to investigate the existence of boundary conditions with asymptotic symmetries given by (\ref{virU1}).

\item Our derivation of the Bekenstein-Hawking entropy via \eqref{SWCFT} has two important weaknesses. First we have not derived the vacuum value \eqref{L0vac}, nor identified the corresponding geometry. This is tied to our lack of knowledge of the classical phase space, which is a persistent shortcoming of works related to the Kerr/CFT correspondence. 

Second, the derivation of \eqref{scan2}, as outlined  in App.\,\ref{app:modwcft}, is strictly valid in the high temperature regime. In particular it assumes that $c$ and $k$ are fixed, and $T_{L,R}\gg1$. In the gravitational side, the temperatures are constrained  via $T_{L}^2-T_R^2=1/4\pi^2$, which in particular allows them to have small values as the black hole reaches extremality. This is somewhat similar to the non-trivial conditions one needs to impose on a CFT$_2$ such that entropy of the BTZ black hole is fully captured \cite{Hartman:2014oaa}. Understanding the conditions on WCFTs that accommodate for the thermodynamic regime of the black hole  remains an open problem worth exploring. 


\end{enumerate}

\section*{Acknowledgements}

We thank Geoffrey Comp\`ere, Laura Donnay, Tom Hartman, Diego Hofman, Andrea Puhm, Andrew Strominger and C\'eline Zwikel for useful discussions.   We also thank the hospitality of the Erwin Schr\"{o}dinger Institute (ESI) in Vienna in March 2019 during the program `Higher Spins and Holography', where part of this work was completed. 
AC is supported by Nederlandse Organisatie voor Wetenschappelijk Onderzoek (NWO) via a Vidi grant, and by the Delta ITP consortium, a program of the NWO that is funded by the Dutch Ministry of Education, Culture and Science (OCW). 
AA and SD are supported in part by the ARC grant ``Holography, Gauge Theories and Quantum Gravity Building models of quantum black holes'', by IISN -- Belgium (convention 4.4503.15) and benefited from the support of the Solvay Family. 
SD is a Research Associate of the Fonds de la Recherche Scientifique F.R.S.-FNRS (Belgium).

\appendix

\section{Aspects of warped conformal symmetries}\label{app:wcft}

In this appendix we record some basic features of warped conformal symmetries, based on \cite{Detournay:2012pc,Castro:2015uaa}. We will focus on the symmetries, conserved charges and modular properties associated to a warped conformal field theory.

Consider a (1+1) dimensional theory defined on a  plane which we describe in terms of two coordinates $(u,v)$. On this plane, we denote as $T(u)$ the operator that generates infinitesimal coordinate transformations in $u$ and $P(u)$ the operator that generates $u$ dependent infinitesimal translations in $v$. The corresponding coordinate transformations are
\be\label{app1}
u ~\to ~ u=f(u')   ~,\quad v~ \to~ v=v' + g(u')~.
\ee

In the quantum theory, $T(u)$ is a right moving energy momentum tensor and $P(u)$ as a right moving $U(1)$ Kac-Moody current.  The associated charges are
\be
L_n =-{i\over 2\pi}\int du\, \zeta_n(u) T(u)~,\quad P_n =-{1\over 2\pi}\int du\, \chi_n(u) P(u)~,
\ee
where we choose the test functions as $\zeta_n=u^{n+1}$ and $\chi_n=u^n$. In terms of the plane charges $(L_n,P_n)$ the commutation relations are
\begin{eqnarray}
\label{eq:canonicalgebra}
[ L_n, L_{n'}] &=&(n-n') L_{n+n'}+\frac{c}{12}n(n^2-1)\delta_{n,-n'}~, \nonumber \\
\, [ L_n, P_{n'}] &=&-n'  P_{n'+n}~, \nonumber \\
\, [ P_n, P_{n'}] &=&k \frac{n}{2}\delta_{n,-n'}~,
\end{eqnarray}
which is a Virasoro-Kac-Moody algebra with central charge $c$ and level $k$. 
The finite transformation properties of the operators are
 \bea\label{app:finite}
P'(u') &=& {\partial u\over \partial u'}\le(P(u)+{k\over 2}{\partial v'\over \partial u} \ri)~,\cr
T'(u') &=& \le({\partial u\over \partial u'}\ri)^2\le(T(u)-{c\over 12}\{u',u\} \ri) + {\partial u\over \partial u'}{\partial v\over \partial u'} P(u) - {k\over 4}\le({\partial v\over \partial u'}\ri)^2~,
\eea
where
\be
\{u',u \}= {{\partial^3 u'\over \partial u^3}\over {\partial u'\over \partial u}} -{3\over 2}\le({{\partial^2 u'\over \partial u^2}\over {\partial u'\over \partial u}}\ri)^2~.
\ee

\subsection{Modular transformations and canonical entropy}\label{app:modwcft}

Consider an arbitrary torus defined by identifications
\be\label{eq:torus1}
(u,v)\sim(u-2\pi \ell, v+2\pi \bar{\ell})\sim(u-2\pi \tau, v+2\pi \bar{ \tau})~.
\ee
The WCFT partition function on this torus is given by 
 \be \label{Z}
 Z_{\bar{\ell}|\ell}(\bar{\tau}|\tau)=\text{Tr}_{\bl|\ell}\bigg(e^{2 \pi i \bt P_0}e^{-2\pi i \t L_0}\bigg)~.
 \ee
where the label $(\bl|\ell)$ denotes the choice of torus. $P_0$ and $L_0$ denote the zero modes of the WCFT generators: 
 \be \label{zero}
 P_0=-{1\over2\pi}{\int_0^{ 2 \pi\ell} P(u)\ du}~,\qquad
 L_0=-{1\over2\pi}{\int_0^{ 2 \pi\ell} T(u)\ du}~.
 \ee

One can do the following warped conformal transformation
\be\label{eq:torus2}
\hat{u}={u\over\ell}~,\qquad\hat{v}=v+{\bl\over\ell} u~,
\ee
 which brings the generic torus in \eqref{eq:torus1} to a canonical torus, i.e. it sets $\bl=0$ and $\ell=1$. This transformation affects also the zero modes, as dictated by \eqref{app:finite}, which allows us to relate the partition function on \eqref{eq:torus1} and \eqref{eq:torus2} as
\be \label{Zcan}
Z_{\bar{\ell}|\ell}(\bar{\tau}|\tau)=e^{\pi i k \bl\big(\bt-{\bl \t \over2\ell}\big)}\hat{Z}\big(\bt-{\bl\t\over\ell }|{\t\over\ell}\big)~,
\ee
where 
\be
\hat{Z}\big(\hat{\bt}|\hat{\t}\big):=Z_{0|1}\big(\hat{\bt}|\hat{\t}\big)=\text{Tr}_{0|1}\bigg(e^{2 \pi i \hat{\bt} \hat P_0}e^{-2\pi i \hat{\t} \hat{ L}_0}\bigg)~.
\ee


The statement that a WCFT is invariant under an $S$-transformation implies 
\be
Z_{0|1}(\bt|\t)=Z_{\bt|\t}(0|-1)~.
\ee
This allows us to rewrite \eqref{Zcan} as
\be
\hat{Z}(z|t)=e^{\pi i k {z^2\over 2t}}\hat{Z}\bigg({z\over t}\bigg|{-1\over t}\bigg)~.
\ee
Thus, a generic partition function can be written as 
\be \label{Zmod}
Z_{\bar{\ell}|\ell}(\bar{\tau}|\tau)=e^{\pi i k \bl\big(\bt-{\t\bl\over2\ell}\big)+\pi i k {z^2\over 2t}}\hat{Z}\bigg({z\over t}\bigg|{-1\over t}\bigg)=e^{\pi i k {\ell\bt^2\over 2\t}}\hat{Z}\bigg({z\over t}\bigg|{-1\over t}\bigg)~,
\ee
where 
\be
z=\bt-{\bl\t\over\ell}~, ~ \ t={\t\over \ell}~.
\ee

The entropy is defined by 
\be\label{app:ent}
S=(1-\t\p_\t-\bt\p_{\bt})\log Z~.
\ee
We are interested in extracting the entropy in the high temperature regime, i.e. in the limit $t\rightarrow -i0$. If $z$ is purely imaginary and the spectrum of $L_0$ is bounded below one can use \eqref{Zmod} to obtain the projected partition function
\be \label{Z projected}
Z\rightarrow e^{\pi i k {\ell\bt^2\over 2\t}}e^{2 \pi i {z\over t} \hat P_0^{\rm vac}}e^{{2\pi i\over t}  \hat{ L}_0^{\rm vac}}~.
\ee
In other words, we can approximate the partition function by the vacuum state on the canonical torus.
Using \eqref{Z projected} in \eqref{app:ent}, the entropy in this regime is
\be \label{scan}
S_{\bl|\ell}(\bt|\t)=2\pi i {z\over t}\hat{P}_0^{\rm vac}+{4 \pi i\over t} \hat{L}_0^{\rm vac}
 \ee
with
\be
z=\bt-{\bl\t\over\ell}, ~ \ t={\t\over \ell}.
\ee
Here $\hat{P}_0^{\rm vac}, \hat{L}_0^{\rm vac}$ are the vacuum values of the zero modes on the canonical torus.

\section{Cardy growth revisited}\label{app:cardy}

In this appendix we revisit the derivation of the entropy in the canonical ensemble for a CFT$_2$. The goal is to set the derivation in the same language as done in Appendix \ref{app:modwcft} for a WCFT.

Consider a 2d CFT on a torus with symmetries $u \rightarrow f(u)$ and $v \rightarrow f(v)$ defined by the following identifications:
\be\label{eq:torus21}
(u,v)\sim(u-2\pi \ell, v+2\pi \bar{\ell})\sim(u-2\pi \tau, v+2\pi \bar{ \tau})~.
\ee
The partition function is written
\be
 Z_{\bl|\ell}(\bt|\t) = \text{Tr}_{\bl|\ell}\bigg(e^{2 \pi i \bt \bar L_0}e^{-2\pi i \t L_0}\bigg)~.
\ee
Using the transformation $u'= \lambda^+ u$ and $v'= \lambda^+ v$, one can map the theory on a canonical circle $(\bl, \ell) = (1,1)$, which implies
\be
Z_{\bl|\ell}(\bt|\t) =Z_{1|1}(\frac{\bt}{\bl}|\frac{\t}{\ell})~.
\ee 
The modular S-transformation in 2d CFTs is 
\be
Z_{1|1}(\bt|\t) = Z_{1|1}(-\frac{1}{\bt}|-\frac{1}{\t})~.
\ee
From this, one gets  
\be
Z_{\bl|\ell}(\bt|\t) = Z_{1|1}(-\frac{\bl}{\bt}|-\frac{\ell}{\t})~.
\ee
At small temperatures, the right hand side can be projected on the vacuum state on the canonical torus, whose charges are denoted by $L_0^{\rm vac}$ and $\bar L_0^{\rm vac}$, to obtain
\be
 \log Z_{\bl|\ell}(\bt|\t) \approx -2 \pi i \frac{\bl}{\bt}  \bar L_0^{\rm vac}+ 2 \pi i \frac{\ell}{\t} L_0^{\rm vac}  ~,
\ee
from which one gets the entropy
\be \label{SCFT}
S_{\bl|\ell}(\bt|\t) \approx - 4 \pi i \frac{\bl}{\bt}  \bar L_0^{\rm vac} +4 \pi i \frac{\ell}{\t} L_0^{\rm vac} ~. 
\ee
This is the well known Cardy formula in the canonical ensemble, which we are recasting for a general torus.

For the Kerr black hole, the identifications that define the Euclidean geometry are 
\be \label{KerrIdCFT}
 (t,\phi) \sim (t,\phi - 2 \pi) \sim (t + i \beta_K, \phi + \theta_K)~.
\ee
From \eqref{eq:xx1}, we identify $u \leftrightarrow t^+$ and $v \leftrightarrow t^-$, and so  
\be
\ell = 2 \pi T_R~, \quad \bl = 2 \pi T_L~, \qquad \tau = -i~, \quad \bt = i~.
\ee
Using these values \eqref{SCFT} together with $L_0^{\rm vac} =  \bar L_0^{\rm vac} = -\frac{c}{24}$,  leads to 
\be   
S = \frac{\pi^2}{3} c (T_L + T_R)~,
\ee
which is the Cardy formula reported in \eqref{CardyEntropy}.

\section{Hidden warped symmetries of the Klein-Gordon operator}\label{app:wave}

In this appendix  we will revisit the hidden conformal symmetries of \cite{Castro:2010fd}: we will show how a mild modification to the ``near region'' allows for an additional term in the wave equation. This addition is naturally interpreted in terms of a hidden warped symmetry.

We will define the ``near region'' as region in phase space where we will impose   
\be\label{eq:limit}
r\omega \ll 1 ~,\qquad M \omega  \ll 1~.
\ee 
In the regime \eqref{eq:limit}, the wave equation \eqref{bb:10} leads to
\bea\label{eq:modwv}
\Bigg[
\partial_r (r-r_-)(r-r_+) \partial_r
&+&{(\omega-\Omega_+ m )^2\over 4\kappa_+^2}{(r_+-r_-) \over (r-r_+)} \cr &&\cr & -&{(\omega-\Omega_- m )^2\over 4\kappa_-^2}{ (r_+-r_-)\over (r-r_-)} +4M^2\omega^2 \Bigg] R(r) =K_\ell  R(r)~ ,
\eea
Relative to the limit in \cite{Castro:2010fd}, where the term last in the bracket is absent, we are arguing here
that the terms
\be
{(\omega-\Omega_\pm m )^2\over 4\kappa_\pm^2}{(r_+-r_-) \over (r-r_{\pm})} ~, \qquad M^2 \omega^2~,
\ee
are the leading and subleading contributions in frequency space, while $r\omega$ contributions in \eqref{bb:10} are negligible relative to these terms. Note that $K_\ell$ also receives $\omega^2$ corrections which should be accounted for too. In terms of analyticity of the wave equation, we are keeping the data of the two regular singular points at $r_{\pm}$ and the constant term in the effective potential; this transforms the irregular singular point at $r\to \infty$ into a regular singularity. 
 
Although we have one additional term in this regime, we can write the wave  equation \eqref{eq:modwv} as
\be\label{eq:wave1}
\left({\cal H}^2 -  \bar{H_0}^2 \right)R(r)=K_\ell R(r)~,
\ee
where ${\cal H}^2$ is the quadratic Casimir of an $sl(2,\RR)$ algebra
\be
{\cal H}^2=-H_0^2+{1\over 2}(H_1H_{-1}+H_{-1}H_1)~,
\ee
and the generators are
\bea\label{eq:v1}
H_1&=&ie^{-2\pi T_R \phi}\left(\Delta^{1/2}\partial_r+{1\over 2\pi T_R}{r-M\over \Delta^{1/2} }\partial_\phi+{2T_L\over T_R}{Mr-a^2\over \Delta^{1/2}}\partial_t\right)~,\cr   H_0&=&{i \over 2\pi T_R}\partial_\phi +2iM{T_L \over T_R}\partial_t~,\cr
H_{-1}&=&ie^{2\pi T_R \phi}\left(-\Delta^{1/2}\partial_r+{1\over 2\pi T_R}{r-M\over \Delta^{1/2} }\partial_\phi+{2T_L\over T_R}{Mr-a^2\over \Delta^{1/2}}\partial_t\right)~, 
\eea
The additional generator in \eqref{eq:wave1} is just a $u(1)$ generator defined as
\bea\label{eq:v23}
   \bar H_0&=&-2iM\partial_t~, 
\eea
which encodes the additional $4M^2\omega^2$ term in \eqref{eq:modwv}.

The interpretation of \eqref{eq:wave1} is interesting. For fixed $K_\ell$,  the solutions can be organized as representations of $sl(2)\times u(1)$, which is compatible with the global isometries of a WCFT. Moreover the wave equations of the form \eqref{eq:wave1} can be identified as those that arise from thermal Warped AdS$_3$ geometries; see for example section 5 of \cite{Song:2017czq}, where the WAdS/WCFT Green's function are discussed.

\bibliographystyle{JHEP-2}
\bibliography{refs-hcs}

\providecommand{\href}[2]{#2}\begingroup\raggedright\begin{thebibliography}{10}

\bibitem{Strominger:1996sh}
A.~Strominger and C.~Vafa, {\it Microscopic origin of the
  {B}ekenstein-{H}awking entropy},  {\em Phys. Lett.} {\bf B379} (1996) 99--104
  [\href{http://arXiv.org/abs/hep-th/9601029}{{\tt hep-th/9601029}}].

\bibitem{Sen:2014aja}
A.~Sen, {\it {Microscopic and Macroscopic Entropy of Extremal Black Holes in
  String Theory}},  {\em Gen. Rel. Grav.} {\bf 46} (2014) 1711
  [\href{http://arXiv.org/abs/1402.0109}{{\tt 1402.0109}}].

\bibitem{DeHaro:2019gno}
S.~De~Haro, J.~van Dongen, M.~Visser and J.~Butterfield, {\it {Conceptual
  Analysis of Black Hole Entropy in String Theory}},
  \href{http://arXiv.org/abs/1904.03232}{{\tt 1904.03232}}.

\bibitem{Strominger:1997eq}
A.~Strominger, {\it Black hole entropy from near-horizon microstates},  {\em
  JHEP} {\bf 02} (1998) 009 [\href{http://arXiv.org/abs/hep-th/9712251}{{\tt
  hep-th/9712251}}].

\bibitem{Maldacena:1998bw}
J.~M. Maldacena and A.~Strominger, {\it Ads(3) black holes and a stringy
  exclusion principle},  {\em JHEP} {\bf 12} (1998) 005
  [\href{http://arXiv.org/abs/hep-th/9804085}{{\tt hep-th/9804085}}].

\bibitem{Brown:1986nw}
J.~D. Brown and M.~Henneaux, {\it {Central Charges in the Canonical Realization
  of Asymptotic Symmetries: An Example from Three-Dimensional Gravity}},  {\em
  Commun.Math.Phys.} {\bf 104} (1986) 207--226.

\bibitem{BTZ}
M.~Banados, C.~Teitelboim and J.~Zanelli, {\it The black hole in
  three-dimensional space-time},  {\em Phys. Rev. Lett.} {\bf 69} (1992)
  1849--1851 [\href{http://arXiv.org/abs/hep-th/9204099}{{\tt
  hep-th/9204099}}].

\bibitem{BHTZ}
M.~Banados, M.~Henneaux, C.~Teitelboim and J.~Zanelli, {\it Geometry of the
  (2+1) black hole},  {\em Phys. Rev.} {\bf D48} (1993) 1506--1525
  [\href{http://arXiv.org/abs/gr-qc/9302012}{{\tt gr-qc/9302012}}].

\bibitem{Maldacena:1997ih}
J.~M. Maldacena and A.~Strominger, {\it {Universal low-energy dynamics for
  rotating black holes}},  {\em Phys. Rev.} {\bf D56} (1997) 4975--4983
  [\href{http://arXiv.org/abs/hep-th/9702015}{{\tt hep-th/9702015}}].

\bibitem{Cvetic:1997xv}
M.~Cvetic and F.~Larsen, {\it {Grey body factors for rotating black holes in
  four-dimensions}},  {\em Nucl. Phys.} {\bf B506} (1997) 107--120
  [\href{http://arXiv.org/abs/hep-th/9706071}{{\tt hep-th/9706071}}].

\bibitem{Cvetic:1997ap}
M.~Cvetic and F.~Larsen, {\it {Greybody factors for black holes in
  four-dimensions: Particles with spin}},  {\em Phys. Rev.} {\bf D57} (1998)
  6297--6310 [\href{http://arXiv.org/abs/hep-th/9712118}{{\tt
  hep-th/9712118}}].

\bibitem{Castro:2010fd}
A.~Castro, A.~Maloney and A.~Strominger, {\it {Hidden Conformal Symmetry of the
  Kerr Black Hole}},  {\em Phys.Rev.} {\bf D82} (2010) 024008
  [\href{http://arXiv.org/abs/1004.0996}{{\tt 1004.0996}}].

\bibitem{Haco:2018ske}
S.~Haco, S.~W. Hawking, M.~J. Perry and A.~Strominger, {\it {Black Hole Entropy
  and Soft Hair}},  {\em JHEP} {\bf 12} (2018) 098
  [\href{http://arXiv.org/abs/1810.01847}{{\tt 1810.01847}}].

\bibitem{Hofman:2011zj}
D.~M. Hofman and A.~Strominger, {\it {Chiral Scale and Conformal Invariance in
  2D Quantum Field Theory}},  {\em Phys.Rev.Lett.} {\bf 107} (2011) 161601
  [\href{http://arXiv.org/abs/1107.2917}{{\tt 1107.2917}}].

\bibitem{Detournay:2012pc}
S.~Detournay, T.~Hartman and D.~M. Hofman, {\it {Warped Conformal Field
  Theory}},  {\em Phys.Rev.} {\bf D86} (2012) 124018
  [\href{http://arXiv.org/abs/1210.0539}{{\tt 1210.0539}}].

\bibitem{Hofman:2014loa}
D.~M. Hofman and B.~Rollier, {\it {Warped Conformal Field Theory as Lower Spin
  Gravity}},  \href{http://arXiv.org/abs/1411.0672}{{\tt 1411.0672}}.

\bibitem{Castro:2015uaa}
A.~Castro, D.~M. Hofman and G.~Sarosi, {\it {Warped Weyl fermion partition
  functions}},  {\em JHEP} {\bf 11} (2015) 129
  [\href{http://arXiv.org/abs/1508.06302}{{\tt 1508.06302}}].

\bibitem{Detournay:2015ysa}
S.~Detournay and C.~Zwikel, {\it {Phase transitions in warped AdS$_{3}$
  gravity}},  {\em JHEP} {\bf 05} (2015) 074
  [\href{http://arXiv.org/abs/1504.00827}{{\tt 1504.00827}}].

\bibitem{Castro:2015csg}
A.~Castro, D.~M. Hofman and N.~Iqbal, {\it {Entanglement Entropy in Warped
  Conformal Field Theories}},  {\em JHEP} {\bf 02} (2016) 033
  [\href{http://arXiv.org/abs/1511.00707}{{\tt 1511.00707}}].

\bibitem{Song:2016gtd}
W.~Song, Q.~Wen and J.~Xu, {\it {Modifications to Holographic Entanglement
  Entropy in Warped CFT}},  {\em JHEP} {\bf 02} (2017) 067
  [\href{http://arXiv.org/abs/1610.00727}{{\tt 1610.00727}}].

\bibitem{Song:2016pwx}
W.~Song, Q.~Wen and J.~Xu, {\it {Generalized Gravitational Entropy for Warped
  Anti-de Sitter Space}},  {\em Phys. Rev. Lett.} {\bf 117} (2016), no.~1
  011602 [\href{http://arXiv.org/abs/1601.02634}{{\tt 1601.02634}}].

\bibitem{Song:2017czq}
W.~Song and J.~Xu, {\it {Correlation Functions of Warped CFT}},  {\em JHEP}
  {\bf 04} (2018) 067 [\href{http://arXiv.org/abs/1706.07621}{{\tt
  1706.07621}}].

\bibitem{Jensen:2017tnb}
K.~Jensen, {\it {Locality and anomalies in warped conformal field theory}},
  {\em JHEP} {\bf 12} (2017) 111 [\href{http://arXiv.org/abs/1710.11626}{{\tt
  1710.11626}}].

\bibitem{Apolo:2018eky}
L.~Apolo and W.~Song, {\it {Bootstrapping holographic warped CFTs or: how I
  learned to stop worrying and tolerate negative norms}},  {\em JHEP} {\bf 07}
  (2018) 112 [\href{http://arXiv.org/abs/1804.10525}{{\tt 1804.10525}}].

\bibitem{Apolo:2018oqv}
L.~Apolo, S.~He, W.~Song, J.~Xu and J.~Zheng, {\it {Entanglement and chaos in
  warped conformal field theories}},  {\em JHEP} {\bf 04} (2019) 009
  [\href{http://arXiv.org/abs/1812.10456}{{\tt 1812.10456}}].

\bibitem{Song:2019txa}
W.~Song and J.~Xu, {\it {Structure Constants from Modularity in Warped CFT}},
  \href{http://arXiv.org/abs/1903.01346}{{\tt 1903.01346}}.

\bibitem{Chen:2019xpb}
B.~Chen, P.-X. Hao and W.~Song, {\it {R\'enyi Mutual Information in Holographic
  Warped CFTs}},  \href{http://arXiv.org/abs/1904.01876}{{\tt 1904.01876}}.

\bibitem{Israel:2004vv}
D.~Isra{\"e}l, C.~Kounnas, D.~Orlando and P.~M. Petropoulos, {\it Electric /
  magnetic deformations of ${S}^3$ and ${A}d{S}_3$, and geometric cosets},
  {\em Fortsch. Phys.} {\bf 53} (2005) 73--104
  [\href{http://arXiv.org/abs/hep-th/0405213}{{\tt hep-th/0405213}}].

\bibitem{Rooman:1998xf}
M.~Rooman and P.~Spindel, {\it Goedel metric as a squashed anti-de {S}itter
  geometry},  {\em Class. Quant. Grav.} {\bf 15} (1998) 3241--3249
  [\href{http://arXiv.org/abs/gr-qc/9804027}{{\tt gr-qc/9804027}}].

\bibitem{Moussa:2003fc}
K.~A. Moussa, G.~Clement and C.~Leygnac, {\it {The Black holes of topologically
  massive gravity}},  {\em Class.Quant.Grav.} {\bf 20} (2003) L277--L283
  [\href{http://arXiv.org/abs/gr-qc/0303042}{{\tt gr-qc/0303042}}].

\bibitem{Anninos:2008fx}
D.~Anninos, W.~Li, M.~Padi, W.~Song and A.~Strominger, {\it {Warped AdS(3)
  Black Holes}},  {\em JHEP} {\bf 0903} (2009) 130
  [\href{http://arXiv.org/abs/0807.3040}{{\tt 0807.3040}}].

\bibitem{Compere:2007in}
G.~Comp\`ere and S.~Detournay, {\it {Centrally extended symmetry algebra of
  asymptotically Godel spacetimes}},  {\em JHEP} {\bf 0703} (2007) 098
  [\href{http://arXiv.org/abs/hep-th/0701039}{{\tt hep-th/0701039}}].

\bibitem{Compere:2008cv}
G.~Comp\`ere and S.~Detournay, {\it {Semi-classical central charge in
  topologically massive gravity}},  {\em Class.Quant.Grav.} {\bf 26} (2009)
  012001 [\href{http://arXiv.org/abs/0808.1911}{{\tt 0808.1911}}].

\bibitem{Compere:2009zj}
G.~Comp\`ere and S.~Detournay, {\it {Boundary conditions for spacelike and
  timelike warped AdS$_3$ spaces in topologically massive gravity}},  {\em
  JHEP} {\bf 0908} (2009) 092 [\href{http://arXiv.org/abs/0906.1243}{{\tt
  0906.1243}}].

\bibitem{Anninos:2010pm}
D.~Anninos, G.~Comp\`ere, S.~de~Buyl, S.~Detournay and M.~Guica, {\it {The
  Curious Case of Null Warped Space}},  {\em JHEP} {\bf 1011} (2010) 119
  [\href{http://arXiv.org/abs/1005.4072}{{\tt 1005.4072}}].

\bibitem{Henneaux:2011hv}
M.~Henneaux, C.~Martinez and R.~Troncoso, {\it {Asymptotically warped anti-de
  Sitter spacetimes in topologically massive gravity}},  {\em Phys. Rev.} {\bf
  D84} (2011) 124016 [\href{http://arXiv.org/abs/1108.2841}{{\tt 1108.2841}}].

\bibitem{Blagojevic:2009ek}
M.~Blagojevic and B.~Cvetkovic, {\it {Asymptotic structure of topologically
  massive gravity in spacelike stretched AdS sector}},  {\em JHEP} {\bf 09}
  (2009) 006 [\href{http://arXiv.org/abs/0907.0950}{{\tt 0907.0950}}].

\bibitem{Anninos:2011vd}
D.~Anninos, S.~de~Buyl and S.~Detournay, {\it {Holography For a De Sitter-Esque
  Geometry}},  {\em JHEP} {\bf 05} (2011) 003
  [\href{http://arXiv.org/abs/1102.3178}{{\tt 1102.3178}}].

\bibitem{Donnay:2015iia}
L.~Donnay and G.~Giribet, {\it {Holographic entropy of Warped-AdS$_{3}$ black
  holes}},  {\em JHEP} {\bf 06} (2015) 099
  [\href{http://arXiv.org/abs/1504.05640}{{\tt 1504.05640}}].

\bibitem{Giribet:2015lfa}
G.~Giribet and M.~Tsoukalas, {\it {Warped-AdS3 black holes with scalar halo}},
  {\em Phys. Rev.} {\bf D92} (2015), no.~6 064027
  [\href{http://arXiv.org/abs/1506.05336}{{\tt 1506.05336}}].

\bibitem{Afshar:2015wjm}
H.~Afshar, S.~Detournay, D.~Grumiller and B.~Oblak, {\it {Near-Horizon Geometry
  and Warped Conformal Symmetry}},  {\em JHEP} {\bf 03} (2016) 187
  [\href{http://arXiv.org/abs/1512.08233}{{\tt 1512.08233}}].

\bibitem{Detournay:2016gao}
S.~Detournay, L.-A. Douxchamps, G.~S. Ng and C.~Zwikel, {\it {Warped AdS$_{3}$
  black holes in higher derivative gravity theories}},  {\em JHEP} {\bf 06}
  (2016) 014 [\href{http://arXiv.org/abs/1602.09089}{{\tt 1602.09089}}].

\bibitem{Zwikel:2016smm}
C.~Zwikel, {\it {BTZ Black Holes and Flat Space Cosmologies in Higher
  Derivative Theories}},  {\em Class. Quant. Grav.} {\bf 34} (2017), no.~8
  085003 [\href{http://arXiv.org/abs/1604.02120}{{\tt 1604.02120}}].

\bibitem{Setare:2017nlu}
M.~R. Setare and H.~Adami, {\it {Asymptotically spacelike warped anti-de Sitter
  spacetimes in generalized minimal massive gravity}},  {\em Class. Quant.
  Grav.} {\bf 34} (2017), no.~12 125008
  [\href{http://arXiv.org/abs/1701.00209}{{\tt 1701.00209}}].

\bibitem{Azeyanagi:2018har}
T.~Azeyanagi, S.~Detournay and M.~Riegler, {\it {Warped Black Holes in
  Lower-Spin Gravity}},  {\em Phys. Rev.} {\bf D99} (2019), no.~2 026013
  [\href{http://arXiv.org/abs/1801.07263}{{\tt 1801.07263}}].

\bibitem{Bardeen:1999px}
J.~M. Bardeen and G.~T. Horowitz, {\it {The Extreme Kerr throat geometry: A
  Vacuum analog of AdS(2) x S**2}},  {\em Phys. Rev.} {\bf D60} (1999) 104030
  [\href{http://arXiv.org/abs/hep-th/9905099}{{\tt hep-th/9905099}}].

\bibitem{Guica:2008mu}
M.~Guica, T.~Hartman, W.~Song and A.~Strominger, {\it {The Kerr/CFT
  Correspondence}},  {\em Phys.Rev.} {\bf D80} (2009) 124008
  [\href{http://arXiv.org/abs/0809.4266}{{\tt 0809.4266}}].

\bibitem{Castro:2013kea}
A.~Castro, J.~M. Lapan, A.~Maloney and M.~J. Rodriguez, {\it {Black Hole
  Monodromy and Conformal Field Theory}},  {\em Phys.Rev.} {\bf D88} (2013)
  044003 [\href{http://arXiv.org/abs/1303.0759}{{\tt 1303.0759}}].

\bibitem{Castro:2013lba}
A.~Castro, J.~M. Lapan, A.~Maloney and M.~J. Rodriguez, {\it {Black Hole
  Scattering from Monodromy}},  {\em Class. Quant. Grav.} {\bf 30} (2013)
  165005 [\href{http://arXiv.org/abs/1304.3781}{{\tt 1304.3781}}].

\bibitem{Mano:1996vt}
S.~Mano, H.~Suzuki and E.~Takasugi, {\it {Analytic solutions of the Teukolsky
  equation and their low frequency expansions}},  {\em Prog. Theor. Phys.} {\bf
  95} (1996) 1079--1096 [\href{http://arXiv.org/abs/gr-qc/9603020}{{\tt
  gr-qc/9603020}}].

\bibitem{Mano:1996gn}
S.~Mano and E.~Takasugi, {\it {Analytic solutions of the Teukolsky equation and
  their properties}},  {\em Prog. Theor. Phys.} {\bf 97} (1997) 213--232
  [\href{http://arXiv.org/abs/gr-qc/9611014}{{\tt gr-qc/9611014}}].

\bibitem{Novaes:2014lha}
F.~Novaes and B.~Carneiro~da Cunha, {\it {Isomonodromy, PainlevÃ© transcendents
  and scattering off of black holes}},  {\em JHEP} {\bf 07} (2014) 132
  [\href{http://arXiv.org/abs/1404.5188}{{\tt 1404.5188}}].

\bibitem{daCunha:2015ana}
B.~Carneiro~da Cunha and F.~Novaes, {\it {Kerr Scattering Coefficients via
  Isomonodromy}},  {\em JHEP} {\bf 11} (2015) 144
  [\href{http://arXiv.org/abs/1506.06588}{{\tt 1506.06588}}].

\bibitem{Wald:1999wa}
R.~M. Wald and A.~Zoupas, {\it {A General definition of 'conserved quantities'
  in general relativity and other theories of gravity}},  {\em Phys. Rev.} {\bf
  D61} (2000) 084027 [\href{http://arXiv.org/abs/gr-qc/9911095}{{\tt
  gr-qc/9911095}}].

\bibitem{Compere:2008cw}
G.~Comp\`ere, S.~Detournay and M.~Romo, {\it {Supersymmetric Godel and warped
  black holes in string theory}},  {\em Phys.Rev.} {\bf D78} (2008) 104030
  [\href{http://arXiv.org/abs/0808.1912}{{\tt 0808.1912}}].

\bibitem{Tonni:2010gb}
E.~Tonni, {\it {Warped black holes in 3D general massive gravity}},  {\em JHEP}
  {\bf 08} (2010) 070 [\href{http://arXiv.org/abs/1006.3489}{{\tt 1006.3489}}].

\bibitem{DetournayGuica2013}
S.~Detournay and M.~Guica, {\it {Stringy SchrÃÂ¶dinger truncations}},  {\em
  JHEP} {\bf 08} (2013) 121 [\href{http://arXiv.org/abs/1212.6792}{{\tt
  1212.6792}}].

\bibitem{Compere:2013bya}
G.~Compere, W.~Song and A.~Strominger, {\it {New Boundary Conditions for
  AdS3}},  {\em JHEP} {\bf 05} (2013) 152
  [\href{http://arXiv.org/abs/1303.2662}{{\tt 1303.2662}}].

\bibitem{Carlip:1994gy}
S.~Carlip, {\it {The Statistical mechanics of the (2+1)-dimensional black
  hole}},  {\em Phys. Rev.} {\bf D51} (1995) 632--637
  [\href{http://arXiv.org/abs/gr-qc/9409052}{{\tt gr-qc/9409052}}].

\bibitem{Carlip:1998wz}
S.~Carlip, {\it {Black hole entropy from conformal field theory in any
  dimension}},  {\em Phys. Rev. Lett.} {\bf 82} (1999) 2828--2831
  [\href{http://arXiv.org/abs/hep-th/9812013}{{\tt hep-th/9812013}}].

\bibitem{Carlip:2011vr}
S.~Carlip, {\it {Effective Conformal Descriptions of Black Hole Entropy}},
  {\em Entropy} {\bf 13} (2011) 1355--1379
  [\href{http://arXiv.org/abs/1107.2678}{{\tt 1107.2678}}].

\bibitem{Carlip:2012ff}
S.~Carlip, {\it {Effective Conformal Descriptions of Black Hole Entropy: A
  Review}},  {\em AIP Conf. Proc.} {\bf 1483} (2012), no.~1 54--62
  [\href{http://arXiv.org/abs/1207.1488}{{\tt 1207.1488}}].

\bibitem{Carlip:2017xne}
S.~Carlip, {\it {Black Hole Entropy from Bondi-Metzner-Sachs Symmetry at the
  Horizon}},  {\em Phys. Rev. Lett.} {\bf 120} (2018), no.~10 101301
  [\href{http://arXiv.org/abs/1702.04439}{{\tt 1702.04439}}].

\bibitem{Donnay:2015abr}
L.~Donnay, G.~Giribet, H.~A. Gonzalez and M.~Pino, {\it {Supertranslations and
  Superrotations at the Black Hole Horizon}},  {\em Phys. Rev. Lett.} {\bf 116}
  (2016), no.~9 091101 [\href{http://arXiv.org/abs/1511.08687}{{\tt
  1511.08687}}].

\bibitem{Afshar:2016wfy}
H.~Afshar, S.~Detournay, D.~Grumiller, W.~Merbis, A.~Perez, D.~Tempo and
  R.~Troncoso, {\it {Soft Heisenberg hair on black holes in three dimensions}},
   {\em Phys. Rev.} {\bf D93} (2016), no.~10 101503
  [\href{http://arXiv.org/abs/1603.04824}{{\tt 1603.04824}}].

\bibitem{Donnay:2016ejv}
L.~Donnay, G.~Giribet, H.~A. Gonzalez and M.~Pino, {\it {Extended Symmetries at
  the Black Hole Horizon}},  {\em JHEP} {\bf 09} (2016) 100
  [\href{http://arXiv.org/abs/1607.05703}{{\tt 1607.05703}}].

\bibitem{Afshar:2016kjj}
H.~Afshar, D.~Grumiller, W.~Merbis, A.~Perez, D.~Tempo and R.~Troncoso, {\it
  {Soft hairy horizons in three spacetime dimensions}},  {\em Phys. Rev.} {\bf
  D95} (2017), no.~10 106005 [\href{http://arXiv.org/abs/1611.09783}{{\tt
  1611.09783}}].

\bibitem{Grumiller:2017otl}
D.~Grumiller, P.~Hacker and W.~Merbis, {\it {Soft hairy warped black hole
  entropy}},  {\em JHEP} {\bf 02} (2018) 010
  [\href{http://arXiv.org/abs/1711.07975}{{\tt 1711.07975}}].

\bibitem{Gonzalez:2017sfq}
H.~Gonzalez, D.~Grumiller, W.~Merbis and R.~Wutte, {\it {New entropy formula
  for Kerr black holes}},  {\em EPJ Web Conf.} {\bf 168} (2018) 01009
  [\href{http://arXiv.org/abs/1709.09667}{{\tt 1709.09667}}].

\bibitem{Donnay:2019zif}
L.~Donnay and G.~Giribet, {\it {Cosmological horizons, Noether charges and
  entropy}},  \href{http://arXiv.org/abs/1903.09271}{{\tt 1903.09271}}.

\bibitem{Donnay:2019jiz}
L.~Donnay and C.~Marteau, {\it {Carrollian Physics at the Black Hole Horizon}},
   \href{http://arXiv.org/abs/1903.09654}{{\tt 1903.09654}}.

\bibitem{Grumiller:2019fmp}
D.~Grumiller, A.~Prez, M.~M. Sheikh-Jabbari, R.~Troncoso and C.~Zwikel, {\it
  {Spacetime structure near generic horizons and soft hair}},
  \href{http://arXiv.org/abs/1908.09833}{{\tt 1908.09833}}.

\bibitem{Troessaert:2013fma}
C.~Troessaert, {\it {Enhanced asymptotic symmetry algebra of $AdS$$_{3}$}},
  {\em JHEP} {\bf 08} (2013) 044 [\href{http://arXiv.org/abs/1303.3296}{{\tt
  1303.3296}}].

\bibitem{Avery:2013dja}
S.~G. Avery, R.~R. Poojary and N.~V. Suryanarayana, {\it {An sl(2,$\mathbb{R}$)
  current algebra from $AdS_3$ gravity}},  {\em JHEP} {\bf 01} (2014) 144
  [\href{http://arXiv.org/abs/1304.4252}{{\tt 1304.4252}}].

\bibitem{Grumiller:2016pq}
D.~Grumiller and M.~Riegler, {\it {Most general AdS$_{3}$ boundary
  conditions}},  {\em JHEP} {\bf 10} (2016) 023
  [\href{http://arXiv.org/abs/1608.01308}{{\tt 1608.01308}}].

\bibitem{Perez:2016vqo}
A.~Perez, D.~Tempo and R.~Troncoso, {\it {Boundary conditions for General
  Relativity on AdS$_{3}$ and the KdV hierarchy}},  {\em JHEP} {\bf 06} (2016)
  103 [\href{http://arXiv.org/abs/1605.04490}{{\tt 1605.04490}}].

\bibitem{Hartman:2014oaa}
T.~Hartman, C.~A. Keller and B.~Stoica, {\it {Universal Spectrum of 2d
  Conformal Field Theory in the Large c Limit}},  {\em JHEP} {\bf 1409} (2014)
  118 [\href{http://arXiv.org/abs/1405.5137}{{\tt 1405.5137}}].

\end{thebibliography}\endgroup

\end{document}